\PassOptionsToPackage{usenames,rgb}{xcolor}
\documentclass[a4paper,12pt]{article}
\pdfoutput=1 
\usepackage[T1]{fontenc}
\usepackage{amsfonts}
\usepackage{ae,aecompl}
\usepackage{amsmath}
\usepackage{amssymb}
\usepackage{amsthm}
\usepackage[pdftex]{graphicx}
\usepackage{multicol,multirow}       
\usepackage[small]{caption}   
\usepackage{enumerate} 
\usepackage{url}
\usepackage{mathdots}
\usepackage{xcolor}
\usepackage{tikz}

\newcommand{\ie}[0]{\textit{i.e.}}

\DeclareGraphicsExtensions{{.jpeg},{.jpg},{.png},{.pdf}}
\graphicspath{{./images/},%
}

\title{Tomographic X-ray data of a lotus root filled with attenuating objects}
\author{T.A.~Bubba\footnote{Department of Physics, Computer Science and Maths, University of Modena and Reggio Emilia, Italy (tatiana.bubba@unimore.it)},
A.~Hauptmann\footnote{Department of Mathematics and Statistics, University of Helsinki, Finland (\mbox{andreas.hauptmann@helsinki.fi})},
S.~Huotari\footnote{Department of Physics, University of Helsinki, Finland (Simo.Huotari@helsinki.fi)}, \\
J.~Rimpel{\"{a}}inen\footnote{Department of Mathematics and Statistics, University of Helsinki, Finland (juho.rimpelainen@helsinki.fi)},
\ and S.~Siltanen\footnote{Department of Mathematics and Statistics, University of Helsinki, Finland (samuli.siltanen@helsinki.fi)}}

\begin{document}

\maketitle

\abstract{\noindent This is the documentation of the tomographic X-ray data of a lotus root, filled with four different attenuating objects, of different sizes. Data are available at {\url{www.fips.fi/dataset.php}}, and can be freely used for scientific purposes with appropriate references to them, and to this document in \url{http://arxiv.org/}{arXiv}. The data set consists of (1) the X-ray sinogram of a single 2D slice of the lotus root with two different resolutions and (2) the corresponding measurement matrices modeling the linear operation of the X-ray transform. Each of these sinograms was obtained from a measured 360-projection fan-beam sinogram by down-sampling and taking logarithms. The original (measured) sinogram is also provided in its original form and resolution. }

\section{Introduction}

The main idea behind the project is to create real CT measurement data for testing sparse-data tomography algorithms. The lotus root has the texture of a potato, presents many holes of different sizes and it is mainly composed of starch (sugar). The structure of the lotus root makes it ideally suitable to stuff it with other objects. In particular, the following four objects, each placed in a different hole of the lotus root, have been chosen: a pencil (whose kernel is made of carbon, covered by wood), a chalk (made of calcium), three pieces of ceramics (still made of calcium, but their shape is rectangular, whilst the chalk is circular), some match-heads (made of sulphur).
Thus, the lotus root filled with this objects enjoys various structures with different shapes, sizes, contrasts and, most remarkably, attenuations, making it a challenging target for typical sparse-data CT applications.

A video report of the data collection session is available at \url{https://www.youtube.com/watch?v=eWwD_EZuzBI}.

\section{Contents of the data set}

The data set contains the following MATLAB\footnote{MATLAB is a registered trademark of The MathWorks, Inc.} data files:
\begin{itemize}
\item  {\tt Data128.mat},
\item  {\tt Data256.mat},
\item  {\tt FullSizeSinograms.mat} and
\item  {\tt {GroundTruthReconstruction.mat}}.
\end{itemize}
The first two of these files contain CT sinograms and the corresponding measurement matrices with two different resolutions; the data in files {\tt Data128.mat} and {\tt Data256.mat} lead to reconstructions with resolutions $128 \times 128$ and $256 \times 256$, respectively. The data file named {\tt FullSizeSinograms.mat} includes the original (measured) sinograms of $120$ and {$360$} projections, and {\tt GroundTruthReconstruction.mat} contains a high-resolution FBP reconstruction computed from the {$360$}-projection sinogram. Detailed contents of each data can be found below.

\bigskip\noindent
{\tt Data128.mat} contains the following variables:
\begin{enumerate}
\item Sparse matrix {\tt A} of size $51480 \times 16384$; measurement matrix.
\item Matrix {\tt m} of size $429 \times 120$; sinogram (120 projections).
\item Scalar {\tt normA}; norm of the matrix {\tt A}.
\item Scalar {\tt normA\_est}; upper bound for the norm of the matrix {\tt A}.
\end{enumerate}

\bigskip\noindent
{\tt Data256.mat} contains the following variables:
\begin{enumerate}
\item Sparse matrix {\tt A} of size $51480 \times 65536$; measurement matrix.
\item Matrix {\tt m} of size $429 \times 120$; sinogram (120 projections).
\item Scalar {\tt normA}; norm of the matrix {\tt A}.
\item Scalar {\tt normA\_est}; upper bound for the norm of the matrix {\tt A}.
\end{enumerate}

\bigskip\noindent
{\tt FullSizeSinograms.mat} contains the following variables:
\begin{enumerate}
\item Matrix {\tt sinogram120} of size $2221 \times 120$; original (measured) sinogram of 120 projections.
\item Matrix {\tt sinogram360} of size $2221 \times 360$; original (measured) sinogram of 360 projections.
\end{enumerate}

\noindent \textbf{Remark.}
{\itshape The resolutions of the above datasets are designed specifically so that the total variation regularization parameter choice rule published in~\cite{Niinimaki2016} can be applied easily. Also, the reason to include the (upper bounds for) norms of the matrices is the following. Some reconstruction methods require that the norm of the system matrix in equation $Ax=m$ is (at most) one. This can be easily enforced like this in MATLAB:
\begin{verbatim}
A = A/normA;
m = m/normA;
\end{verbatim}
After these lines of code the equation is equivalent to the original but the norm of the system matrix is (at most) one.}

Finally, the dataset called 
{\tt {GroundTruthReconstruction.mat}} contains the following variables:
\begin{enumerate}
\item Matrix {\tt FBP360} of size $1500 \times 1500$; a high-resolution filtered back-projection reconstruction computed from the larger sinogram of 360 projections of the lotus (``ground truth'').
\end{enumerate}
Also, we provide the user with the MATLAB routine producing the filtered back-projection reconstruction (script file {\tt FBPgroundtruthRec.m}). 

\medskip

Details on the X-ray measurements are described in Section \ref{sec:Measurements} below.
The model for the CT problem is
\begin{equation}\label{eqn:Axm}
 {\tt A*x=m(:)},
\end{equation}
where {\tt m(:)} denotes the standard vector form of matrix {\tt m} in MATLAB and {\tt x} is the reconstruction in vector form. In other words, the reconstruction task is to find a vector {\tt x} that (approximately) satisfies \eqref{eqn:Axm} and possibly also meets some additional regularization requirements.
A demonstration of the use of the data is presented in Section \ref{sec:Demo} below.

\section{X-ray measurements}\label{sec:Measurements}

\begin{figure}[t]
\begin{picture}(390,180)
\put(17,0){\includegraphics[width=350pt]{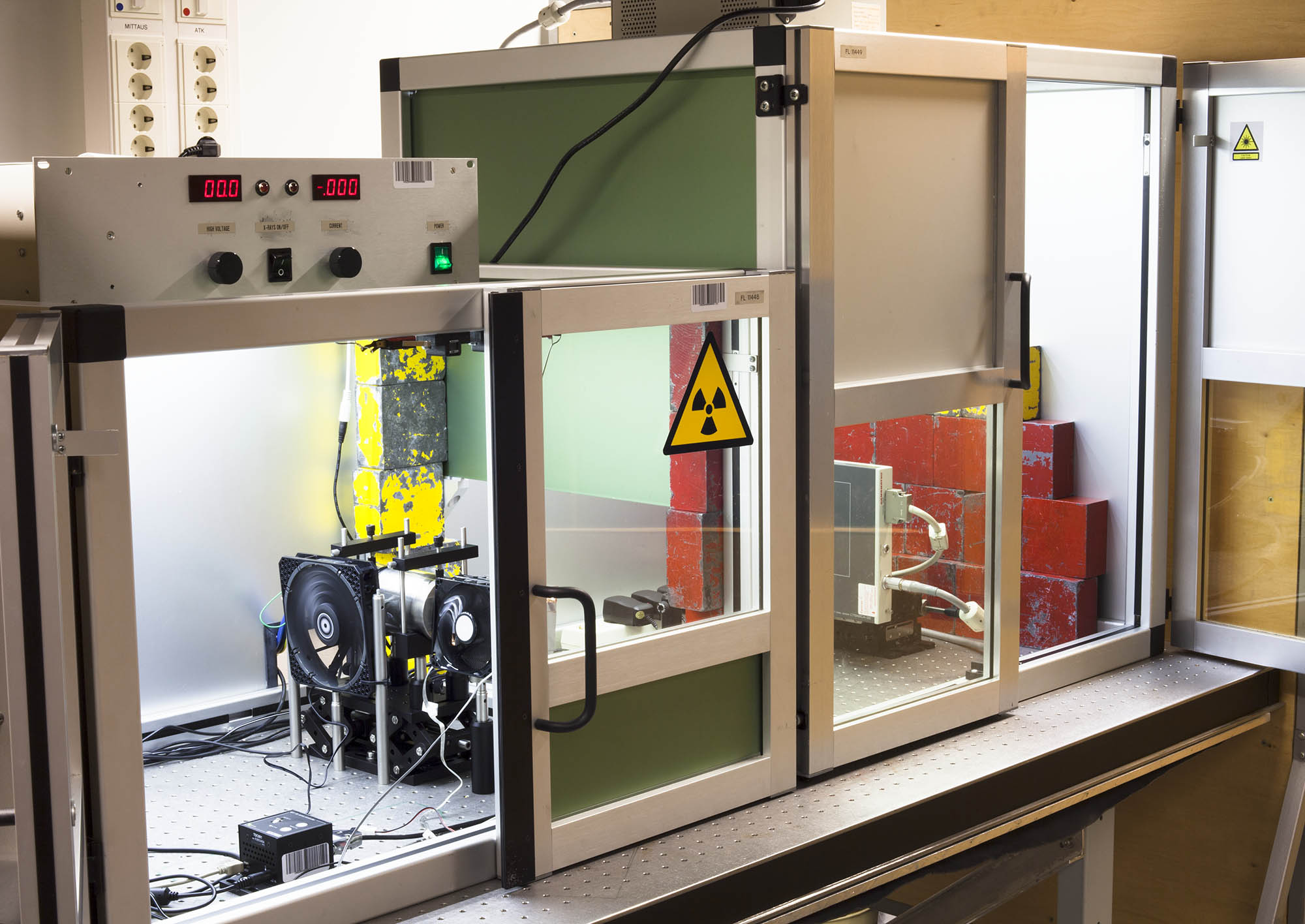}}
\end{picture}
\caption{The custom-made measurement device at University of Helsinki.}\label{fig:Lab}
\end{figure}

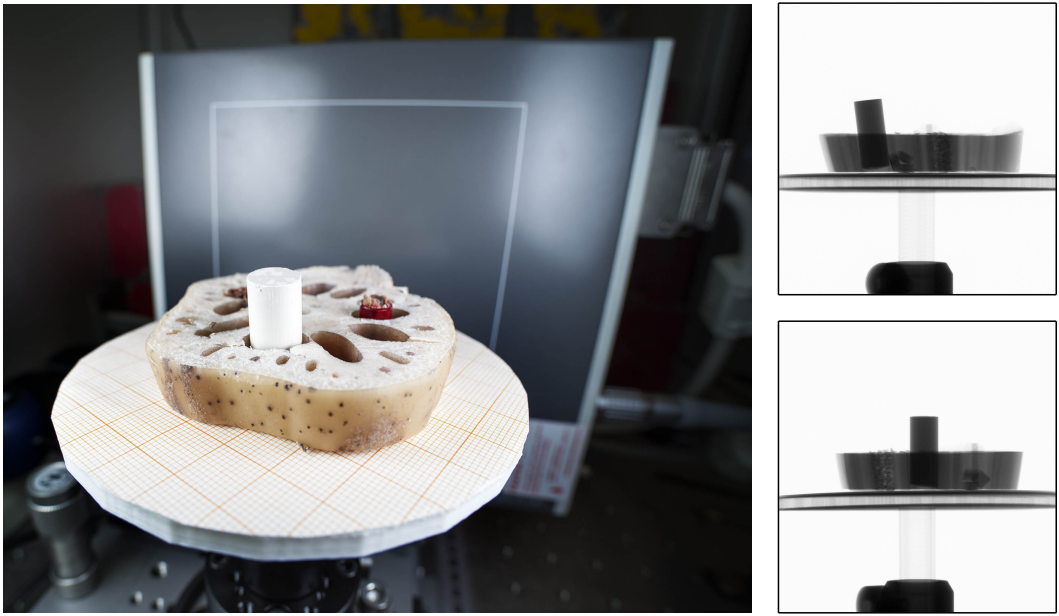
\begin{figure}
\begin{picture}(390,180)
\linethickness{0.2mm}
\put(0,0){\includegraphics[width=280pt]{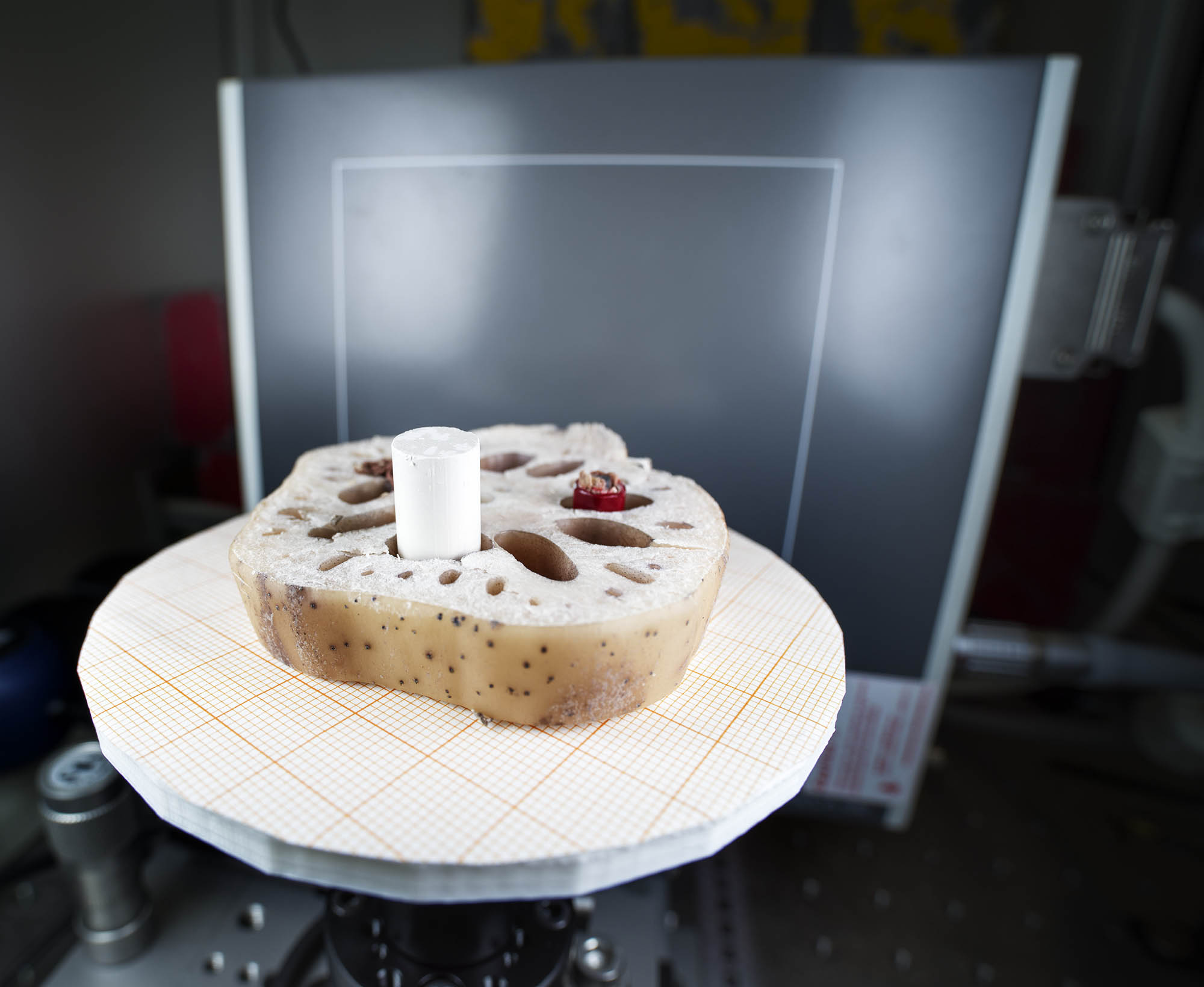}}
\put(290,0){\includegraphics[height=110pt]{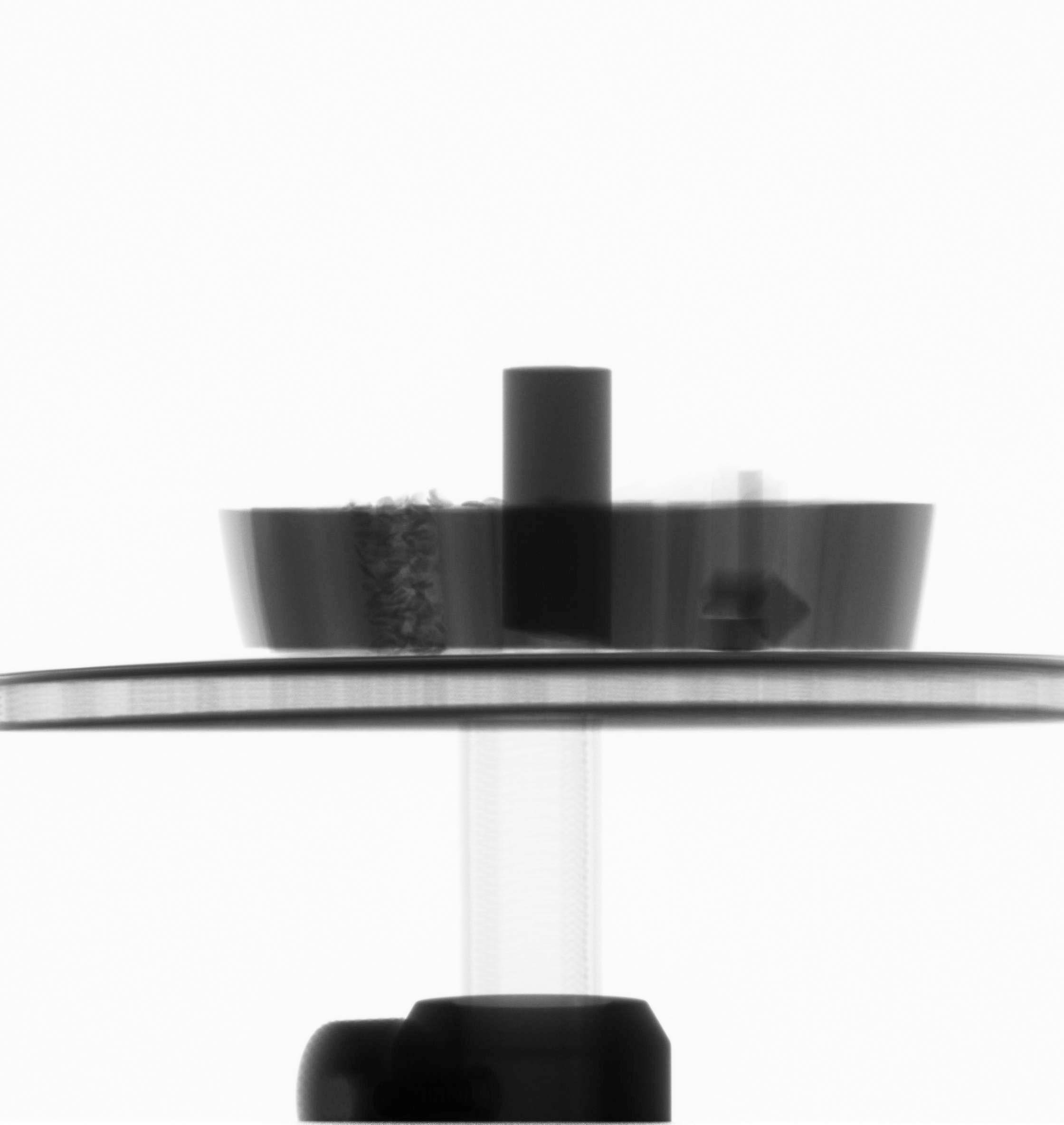}}
\put(290,0){\line(0,1){110}}
\put(290,110){\line(1,0){104}}
\put(394,110){\line(0,-1){110}}
\put(394,0){\line(-1,0){104}}
\put(290,120){\includegraphics[height=110pt]{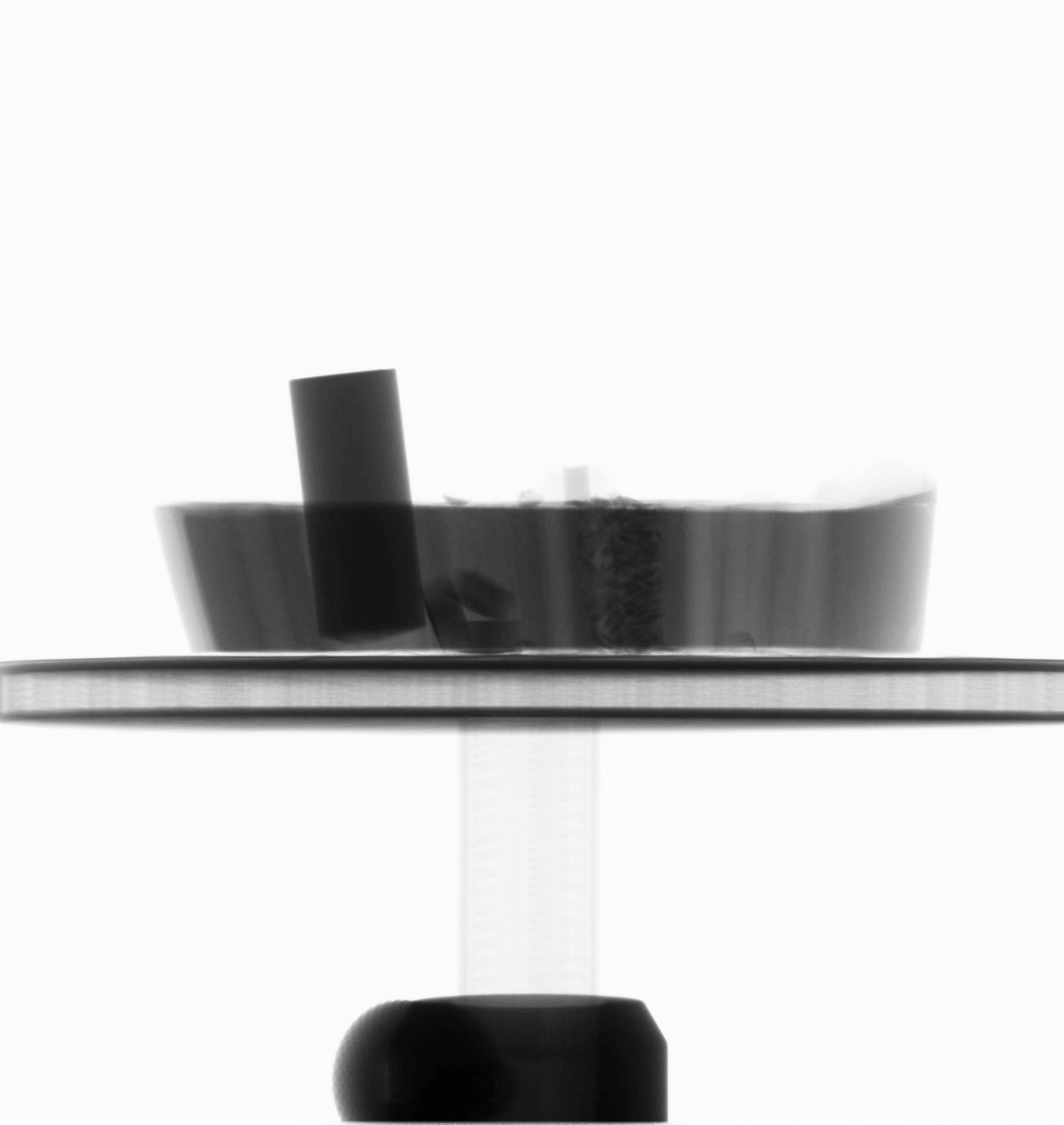}}
\put(290,120){\line(0,1){110}}
\put(290,230){\line(1,0){104}}
\put(394,230){\line(0,-1){110}}
\put(394,120){\line(-1,0){104}}
\end{picture}
\caption{\emph{Left}: Experimental setup used for collecting tomographic X-ray data. The detector plane is behind the lotus root target with the active area indicated by a white square. The target is attached to a computer-controlled rotator platform. \emph{Right}: Two examples of the resulting 2D projection images. The fan-beam data in the sinograms consist of the (down-sampled) central rows of the 2D projection images.}\label{fig:SetupAndProjections}
\end{figure}

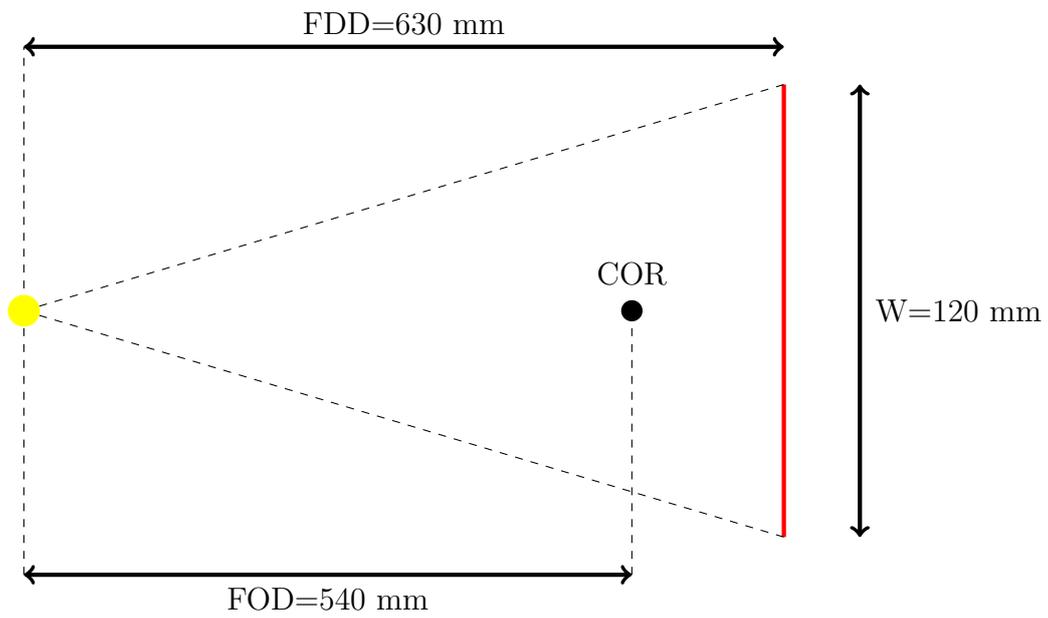
\begin{figure}
\begin{tikzpicture}[scale=2.0]
\draw[ultra thick,red] (2.5,-1.5) -- (2.5,1.5);  
\draw[<->, ultra thick] (-2.5,1.75) -- (2.5,1.75) node[above,midway]{FDD=630 mm}; 
\draw[<->, ultra thick] (-2.5,-1.75) -- (1.5,-1.75) node[below,midway]{FOD=540 mm}; 
\draw[<->, ultra thick] (3.0,-1.5) -- (3.0,1.5); 
\draw[dashed] (-2.5,-1.75) -- (-2.5,1.75); 
\draw[dashed] (1.5,-1.75) -- (1.5,0.0); 
\draw[dashed] (-2.5,0.0) -- (2.5,-1.5);
\draw[dashed] (-2.5,0.0) -- (2.5,1.5); 
\draw (3.65,0.0) node{W=120 mm}; 
\fill[thick] (1.5,0.0) circle (2pt);
\draw (1.5,0.25) node{COR};
\fill[thick, yellow] (-2.5,0.0) circle (3pt);
\end{tikzpicture}
\bigskip
\caption{Geometry of the measurement setup. Here FOD and FDD denote the focus-to-object distance and the focus-to-detector distance, respectively; the black dot COR is the center-of-rotation. The width of the detector (\ie{}, the red thick line) is denoted by W. The yellow dot is the X-ray source. To increase clarity, the $x$-axis and $y$-axis in this image are not in scale.}\label{k1}
\end{figure}

The data  in the sinograms are X-ray tomographic (CT) data of a 2D cross-section of the lotus root measured with a custom-built $\mu$CT device shown in Figure \ref{fig:Lab}. 
\begin{itemize}
\item The X-ray tube is a model XTF5011 manufactured by Oxford Instruments. This
model is no longer sold by Oxford Instruments, although they have newer, similar models available. The tube uses a molybdenum (Z = 42) target. 
\item The rotation stage is a Thorlabs model CR1/M-Z7.
\item The flat panel detector is a Hamamatsu Photonics C7942CA-22. The active area of the flat panel detector is 120 mm $\times$ 120 mm. It consists of a 2400 $\times$ 2400 array of 50 $\mu$m pixels. According to the manufacturer the number of active pixels is 2240 $\times$  2344. However, the image files actually generated by the camera were 2240 $\times$  2368 pixels in size. The device was assembled by Alexander Meaney as a MSc thesis project \cite{Meaney2015}.
\end{itemize}
The measurement arrangement is illustrated in Figure \ref{fig:SetupAndProjections} and the measurement geometry is shown in Figure \ref{k1}.

A  set of $360$ cone-beam projections with resolution $2304 \times 2296$ and angular step of one $(1)$ degree was measured. The exposure time was $1000$ ms (in other words, one second). The X-ray tube acceleration voltage was 50 kV and tube current 1\,mA. See Figure \ref{fig:SetupAndProjections} for two examples of the resulting projection images. 

From the 2D projection images the middle rows corresponding to the central horizontal cross-section of the lotus root were taken to form a fan-beam sinogram of resolution $2221 \times 120$ (variable {\tt sinogram120} in file {\tt FullSizeSinograms.mat}). This sinogram was further down-sampled by binning and taken logarithms of to obtain the sinograms {\tt m} in files {\tt Data128.mat} and {\tt Data256.mat}.

\begin{figure}
\begin{picture}(390,390)

\put(195,390){\color{lightgray}\line(1,-4){90}}
\put(195,390){\color{lightgray}\line(-1,-4){90}}

\put(145,130){\line(1,0){100}}
\put(145,155){\line(1,0){100}}
\put(145,180){\line(1,0){100}}
\put(145,205){\line(1,0){100}}
\put(145,230){\line(1,0){100}}
\put(145,130){\line(0,1){100}}
\put(170,130){\line(0,1){100}}
\put(195,130){\line(0,1){100}}
\put(220,130){\line(0,1){100}}
\put(245,130){\line(0,1){100}}
\put(152,215){$x_1$}
\put(152,190){$x_2$}
\put(155,162){$\vdots$}
\put(152,140){$x_N$}
\put(172,215){$x_{\scriptscriptstyle N+1}$}
\put(172,190){$x_{\scriptscriptstyle N+2}$}
\put(180,162){$\vdots$}
\put(174,140){$x_{2N}$}
\put(200,214){$\cdots$}
\put(200,189){$\cdots$}
\put(200,165){$\ddots$}
\put(200,139){$\cdots$}
\put(225,140){$x_{N^2}$}

\put(105,5){\line(1,0){180}}
\put(105,30){\line(1,0){180}}
\put(105,5){\line(0,1){25}}
\put(130,5){\line(0,1){25}}
\put(155,5){\line(0,1){25}}
\put(260,5){\line(0,1){25}}
\put(285,5){\line(0,1){25}}
\put(110,13){$m_1$}
\put(135,13){$m_2$}
\put(160,14){$\cdots$}
\put(241,14){$\cdots$}
\put(265,13){$m_N$}

\end{picture}
\caption{The organization of the pixels in the sinograms {\tt m}\,=\,$[m_1,m_2,\ldots,m_{120N}]^T$ and reconstructions {\tt x}\,=\,$[x_1,x_2,\ldots,x_{N^2}]^T$ in the data in {\tt Data128.mat} and {\tt Data256.mat} ($N=128$ or $N=256$). The picture shows the organization for the first projection; after that the target takes $3$ degree steps counter-clockwise (or equivalently the source and detector take $3$ degree steps clockwise) and the following columns of {\tt m} are determined in a similar way.}\label{fig:pixelDemo}
\end{figure}
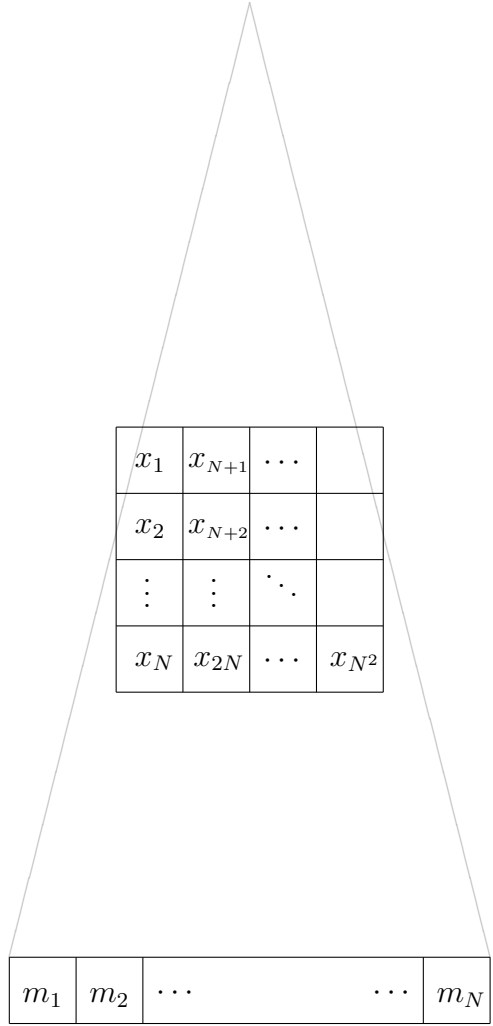

The organization of the pixels in the sinograms and reconstructions is illustrated in Figure \ref{fig:pixelDemo}. The pixel sizes of the reconstructions are $0.627$ mm and $0.315$ mm in the data in {\tt Data128.mat} and {\tt Data256.mat}, respectively.

\begin{figure}
\includegraphics[width=\textwidth]{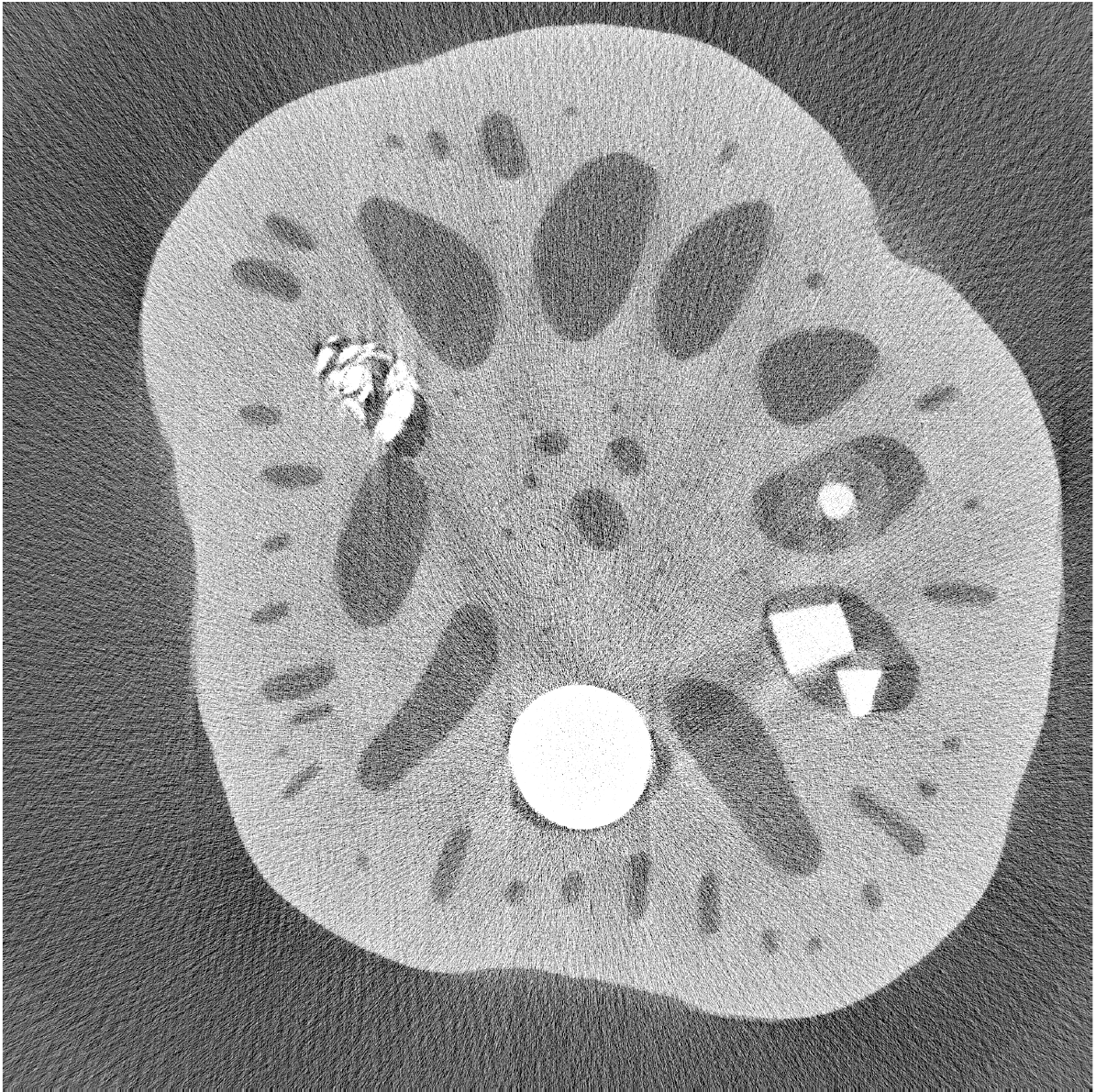}
\caption{The high-resolution filtered back-projection reconstruction {\tt FBP360} of the lotus root computed from 360 projections.}\label{k2}
\end{figure}

In addition, a larger set of $360$ projections of the same lotus root using the same imaging setup and measurement geometry but with a finer angular step of one $(1)$ degrees was measured (variable {\tt sinogram360} in file {\tt FullSizeSinograms.mat}). The high-resolution ground truth reconstruction (variable {\tt FBP360} in file {\tt GroundTruthReconstruction.mat}) was computed from this data using filtered back-projection algorithm, see Figure \ref{k2}.

\clearpage
\section{Example of using the data}\label{sec:Demo}

\begin{figure}
\includegraphics[width=\textwidth]{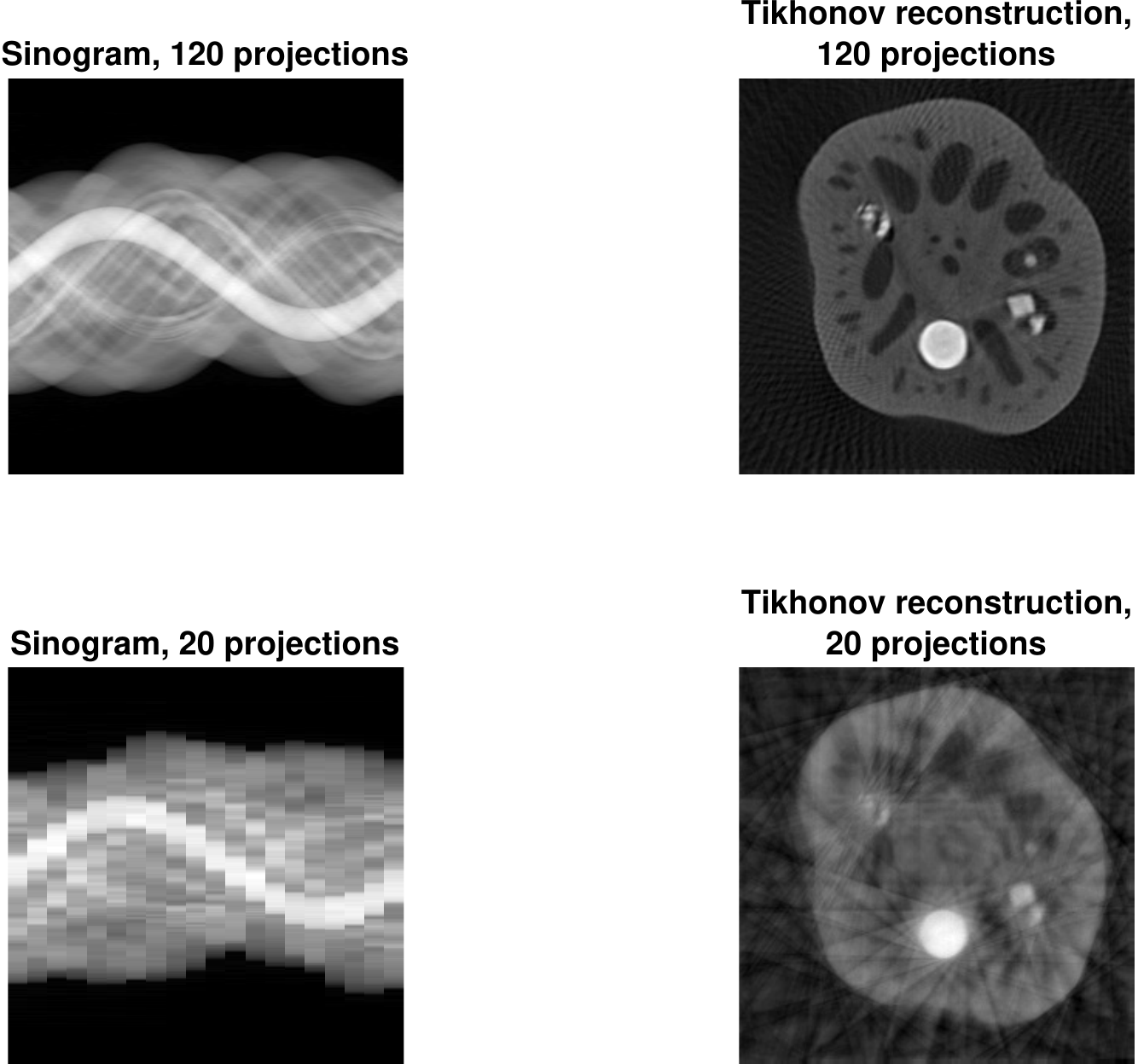}
\bigskip
\caption{First row: sinogram and corresponding Tikhonov regularized reconstruction with 120 projection. Second row: sinogram and corresponding Tikhonov regularized reconstruction with 20 projection.}\label{fig:Tikh}
\end{figure}

The following MATLAB code demonstrates how to use the data. The code is also provided as the separate MATLAB script file {\tt example.m} and it assumes the data files (or in this case at least the file {\tt Data256.mat}) are included in the same directory with the script file. The resulting reconstructed images are reported in Figure \ref{fig:Tikh}.

\begin{verbatim}
% Load the measurement matrix and the sinogram from
% file Data256.mat
load Data256

% Compute a Tikhonov regularized reconstruction using
% conjugate gradient algorithm pcg.m
N     = sqrt(size(A,2));
alpha = 10; % regularization parameter
fun   = @(x) A.'*(A*x)+alpha*x;
b     = A.'*m(:);
x     = pcg(fun,b);

% Compute a Tikhonov regularized reconstruction from only
% 20 projections
[mm,nn] = size(m);
ind     = [];
for iii=1:nn/6
    ind = [ind,(1:mm)+(6*iii-6)*mm];
end
m2    = m(:,1:6:end);
A     = A(ind,:);
alpha = 10; % regularization parameter
fun   = @(x) A.'*(A*x)+alpha*x;
b     = A.'*m2(:);
x2    = pcg(fun,b);

% Take a look at the sinograms and the reconstructions
figure
subplot(2,2,1)
imagesc(m)
colormap gray
axis square
axis off
title('Sinogram, 120 projections')
subplot(2,2,3)
imagesc(m2)
colormap gray
axis square
axis off
title('Sinogram, 20 projections')
subplot(2,2,2)
imagesc(reshape(x,N,N))
colormap gray
axis square
axis off
title({'Tikhonov reconstruction,'; '120 projections'})
subplot(2,2,4)
imagesc(reshape(x2,N,N))
colormap gray
axis square
axis off
title({'Tikhonov reconstruction,'; '20 projections'})
\end{verbatim}

\end{document}